\RequirePackage{filecontents}

\RequirePackage{snapshot}
\documentclass[aps,prl,reprint,showpacs,secnumarabic]{preprint}

\let\Paragraph=\paragraph
\AtBeginDocument{\renewcommand{\paragraph}[1]{\Paragraph{#1}}}
\renewcommand\documentclass[2][]{}
\documentclass[aps,prl,twocolumn,showpacs,floatfix]{revtex4-1}

\usepackage[T1]{fontenc}
\usepackage{amsmath}
\usepackage{txfonts}
\DeclareMathAlphabet{\mathbit}{OT1}{txr}{b}{it}
\usepackage{microtype}
\usepackage{graphicx}
\graphicspath{{figures/}}
\usepackage{color}
\usepackage[breaklinks=true,
pdfauthor={Michael Klaiber, Enderalp Yakaboylu, Heiko Bauke, Karen Z. Hatsagortsyan, and Christoph H. Keitel},
pdftitle={Under-the-barrier dynamics in laser-induced relativistic tunneling}]{hyperref}
\usepackage{braket}

\renewcommand*{\vec}[1]{\ifcat a#1\mathbit{#1}\else\boldsymbol{#1}\fi}
\DeclareMathOperator{\Airy}{Ai}

\newcommand*{\I}{\mathrm{i}}
\newcommand*{\e}{\mathrm{e}}

\makeatletter
\def\paragraph#1{%
  \@startsection
    {paragraph}%
    {4}%
    {\parindent}%
    {\z@}%
    {-1sp}%
    {\normalfont\normalsize\itshape}{#1.---}%
}%
\makeatother

\begin{document}

\title{Under-the-barrier dynamics in laser-induced relativistic tunneling}

\author{Michael Klaiber}
\author{Enderalp Yakaboylu}
\author{Heiko Bauke}
\author{Karen Z. Hatsagortsyan}
\author{Christoph H. Keitel}
\affiliation{Max-Planck-Institut f{\"u}r Kernphysik, Saupfercheckweg~1,
  69117~Heidelberg, Germany} 
\date{\today}

\begin{abstract}
  The tunneling dynamics in relativistic strong-field ionization is
  investigated with the aim to develop an intuitive picture for the
  relativistic tunneling regime. We demonstrate that the tunneling
  picture applies also in the relativistic regime by introducing
  position dependent energy levels. The quantum dynamics in the
  classically forbidden region features two time scales, the typical
  time that characterizes the probability density's decay of the
  ionizing electron under the barrier (Keldysh time) and the time
  interval which the electron spends inside the barrier
  (Eisenbud-Wigner-Smith tunneling time). In the relativistic regime, an
  electron momentum shift as well as a spatial shift along the laser
  propagation direction arise during the under-the-barrier motion which
  are caused by the laser magnetic field induced Lorentz force. The
  momentum shift is proportional to the Keldysh time, while the
  wave-packet's spatial drift is proportional to the
  Eisenbud-Wigner-Smith time. The signature of the momentum shift is
  shown to be present in the ionization spectrum at the detector and,
  therefore, observable experimentally. In contrast, the signature of
  the Eisenbud-Wigner-Smith time delay disappears at far distances for
  pure quasistatic tunneling dynamics.
\end{abstract}


\maketitle


\paragraph{Introduction}

With present-day strong lasers providing intensities of the order of
$10^{22}\,\text{W}/\text{cm\textsuperscript{2}}$ \cite{Yanovsky_2008}
and employing highly charged ions, the relativistic regime of strong
field atomic processes, in particular, strong field ionization of highly
charged ions \cite{Hetzheim_2009,*Bauke_2011}, is accessible
\cite{DiPiazza:etal:2012:high-intensity}. Relativistic generalizations
of the strong field approximation (SFA)
\cite{Reiss_1990,*Reiss_1990b,Keldysh_1965,Faisal_1973,Reiss_1980} and
the quasiclassical theory of ionization
\cite{Popov_1997,*Mur_1998,Perelomov_1967,*Ammosov_1986} provide simple
ways to calculate ionization probabilities in the relativistic regime
\cite{Crawford_1994,*Klaiber_2007,Milosevic_2002a,*Milosevic_2002d}.

The ionization occurs in the so-called tunneling regime when the laser's
electric field amplitude $E_0$ is smaller than those required for
over-the-barrier ionization \cite{Augst:etal:1989:Tunneling_ionization}
and the Keldysh parameter $\gamma=\omega\sqrt{2I_p}/E_0$
\footnote{Atomic units are used throughout the Letter unless stated
  otherwise.}  is well below unity with the ionization potential $I_p$
and the laser's angular frequency $\omega$.  The Keldysh parameter
defines the time scale $\tau_{K}=\gamma/\omega$ during which individual
momentum components of the ionized wave packet are formed
\cite{Keldysh_1965}. When this time scale is shorter than the laser
period, i.\,e., $\gamma\ll 1$, the laser field can be treated as
quasistatic.

In the nonrelativistic as well as in the relativistic case, the electron
travels during its journey from the bound state into the continuum
through a region which is forbidden according to classical mechanics.
For nonrelativistic dynamics, when the typical electron velocities are
negligible compared to the speed of light, it is legitimate to treat the
laser field as a homogeneous time-dependent electric field $\vec E(t)$
and to neglect its magnetic component. In this case a well-known
intuitive picture for the tunneling regime arises in which the electron
tunnels out through the effective potential
$V_{\mathrm{eff}}(\vec{r},t)=V(\vec{r})+\vec{r}\cdot\vec{E}(t)$ formed
by the atomic potential $V(\vec{r})$ and the laser field.  The Keldysh
time $\tau_{K}$ may be interpreted as the time that a classical free
electron would need to cover the length $l\sim I_p/E_0$ of the tunneling
barrier at the characteristic velocity of a bound electron
$\upsilon_a=\sqrt{2I_p}$.  The contribution of the laser's magnetic
field component becomes non-negligible when we enter the relativistic
regime. A transverse laser field with perpendicular electric and
magnetic components cannot be described solely by a scalar
potential. Consequently, a generalization of the tunneling picture into
the relativistic regime is not straightforward \cite{Reiss_2008} and
there is no clear intuitive picture for the relativistic quasistatic
ionization dynamics in the classically forbidden region.  

Later, we will identify the so-called tunneling time $\tau_E$ as a
second time scale that is fundamental for the electron dynamics in the
classically forbidden region. The tunneling time through a potential
barrier (Eisenbud-Wigner-Smith time delay
\cite{Eisenbud_1948,*Wigner_1955,*Smith_1960}) has been a long-standing
problem in quantum mechanics
\cite{MacColl_1932,deCarvalho_2002,Baz_1966,*Peres_1980,*Davies_2005,*Landauer_1994,*Sokolovski_2008,*Ban_2010,*Galapon:2012:Barrier_Traversal_Time}.
Interest to this problem has been renewed recently by experimental
attempts to measure the temporal delay at the electron emergence into
the continuum from a bound state during the laser-induced
(nonrelativistic) tunneling
\cite{Eckle_2008a,*Eckle_2008b,*Pfeiffer_2012}.  A vanishing tunneling
time within the experimental error has been observed in these
experiments.

In order to develop an intuitive picture for the relativistic tunneling
regime of strong field ionization, we carry out a detailed analysis of
the electron dynamics through the classically forbidden region in the
relativistic regime and compare it with the corresponding
nonrelativistic case.  The characteristic time scales are examined.  Our
analysis is based on the SFA
\cite{Reiss_1990,*Reiss_1990b,Keldysh_1965,Faisal_1973,Reiss_1980} as
well as on the exact solution of the relativistic wave equations and
numerical simulations.  The possibility for the experimental detection
of the specific relativistic features of the classically forbidden
dynamics is also elaborated.

\paragraph{System}

A highly charged hydrogenlike ion with nuclear charge $Z\lessapprox100$
at the origin of our coordinate system is considered to interact with a
strong linearly polarized laser field with electric field of amplitude
$E_0$ in $x$-, magnetic field of amplitude $B_0$ in $y$- and propagation
in \mbox{$z$-direction} \mbox{($E_0=B_0$)}.  Position and momenta of
electrons that are ionized by the laser pulse may be measured by a
detector far away from the interaction zone.  Throughout this Letter we
describe the laser field via its vector potential in electric field
gauge \cite{Reiss:1979:gauge,Grynberg:2010:Quantum_Optics}, i.\,e.,
$\phi(\vec{r}, t)=-\vec{r}\cdot \vec{E}(\eta)$ and $\vec{A}(\vec{r},
t)=-\hat{\vec{k}}[\vec{r}\cdot \vec{E}(\eta)]/c$ with the laser phase
$\eta=t-z/c$, the electron's coordinate $\vec{r}=(x,y,z)$, the unit
vector in laser propagation direction $\hat{\vec{k}}$, and the speed of
light $c$.  The dynamics of the electron state $\ket{\Psi}$ in the rest
frame of the ionic core is then defined via the Dirac equation
\begin{equation}
  \I\partial_t\ket{\Psi} = 
  \left\{
    c\boldsymbol{\alpha}\cdot[\hat{\vec{p}}+\vec{A}(\vec{r}, t)] + 
    V(\vec{r}) - \phi(\vec{r}, t) + \beta c^2
  \right\}\ket{\Psi}
  \label{eq:DE}
\end{equation}
with the momentum operator $\hat{\vec{p}}$ and the Dirac matrices
$\boldsymbol{\alpha}$ and $\beta$.  In the following, we use an extreme
but feasible set of parameters, viz.\ $I_p/c^2=0.25$ and $E_0/E_a=1/30$,
with $E_a=Z^3$ denoting the characteristic atomic field strength,
representing a hydrogenlike ion with nuclear charge of $Z\approx 90$ in
the tunneling regime~\cite{Augst:etal:1989:Tunneling_ionization}.

\paragraph{Relativistic strong field approximation}

\begin{figure}[b]
  \hfill
  \includegraphics[scale=0.49]{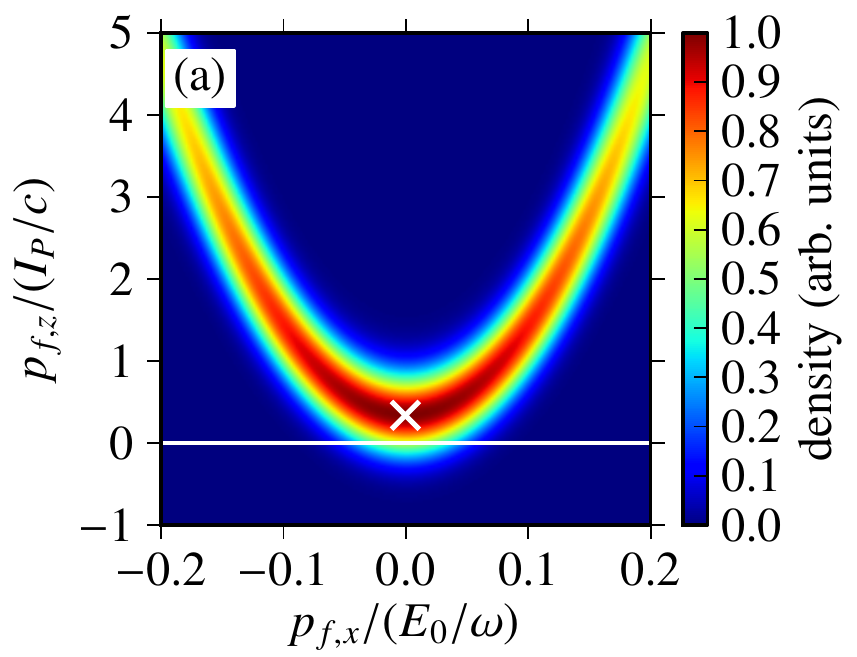}%
  \hfill\hfill
  \includegraphics[scale=0.49]{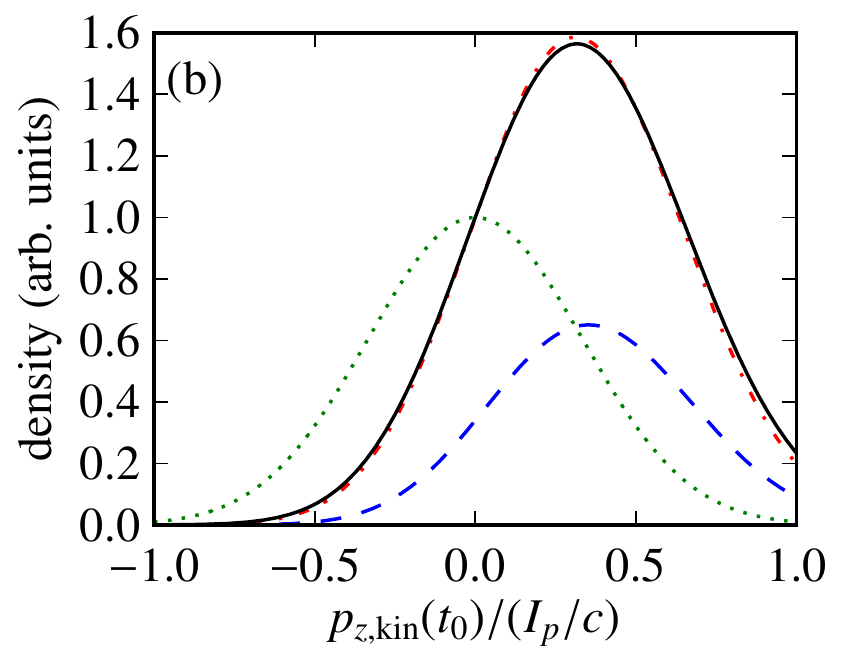}%
  \hfill\mbox{}
  \caption{(a) Density plot of the asymptotic spin-averaged electron
    momentum distribution at the detector. The distribution's maximum
    (white cross) is shifted from $p_{f,z}=0$ (nonrelativistic limit,
    white line) to $p_{f,z}=I_p/(3c)$.  (b) The distribution of the
    electron's kinetic momentum in the laser propagation direction at
    the tunnel exit at time $t_0$, with maximal laser electric field, in
    the electric dipole approximation (green, dotted line), including
    the magnetic dipole correction (blue, dashed line) and the magnetic
    dipole plus the mass-shift correction (red, dot-dashed line) and
    fully relativistically via the Dirac equation in the SFA (black,
    solid line).  The applied parameters are $I_p/c^2=0.25$ and
    $E_0/E_a=1/30$.}
  \label{fig:ted1}
\end{figure}

The SFA allows us to calculate the ionized part of the asymptotic
electron wave packet at a detector in the remote future.  As we are
interested in the under-the-barrier dynamics of the electron, the
momentum distribution at the tunnel exit time $t_0$ is deduced
analytically from the distribution at the detector, assuming free
propagation in the laser field outside the barrier. The asymptotic
ionization amplitude in momentum space reads
\cite{Becker:etal:2002:ATI,*Klaiber:etal:2006:Gauge-invariant_SFA,*Klaiber:etal:2013:unpublished_I,*Klaiber:etal:2013:unpublished_II}
\begin{equation} 
  \braket{\vec{p}_{\!f},s_f | \Psi} = 
  \int_{-\infty}^\infty
  \braket{ 
    \psi_{\vec p_{\!f},s_f} |
    \vec{r}\cdot\vec{E}(\eta) |
    \tilde \psi_0
  }
  \,\mathrm{d}\eta\,,
  \label{eq:a_SFA}
\end{equation}
where $\vec{p}_{\!f}=(p_{f,x},p_{f,y},p_{f,z})$ and $s_f$ denote the
electron's final momentum and spin, respectively, $\ket{\psi_{\vec
    p,s}}$ is the Volkov state \cite{Wolkov:1935:Dirac} with momentum
$\vec{p}$ and spin $s$, while $\ket{\tilde\psi_0}$ is the dressed Dirac
bound state spinor of a short range potential with binding energy $-I_p$
\cite{Becker:etal:2002:ATI,*Klaiber:etal:2006:Gauge-invariant_SFA,*Klaiber:etal:2013:unpublished_I,*Klaiber:etal:2013:unpublished_II}.
The relativistic wave function $\ket{\psi_0}$ is used in the
relativistic as well as in the nonrelativistic calculation allowing to
trace back the features in the under-the-barrier dynamics to the
Hamiltonian's different corrections to the electric dipole
approximation.  Calculating the integral (\ref{eq:a_SFA}) via the saddle
point method, the momentum distributions at the detector, see
Fig.~\ref{fig:ted1}~(a), and at the tunnel exit are derived; see
Fig.~\ref{fig:ted1}~(b). While parabolic wings of size $p_{f,z} \sim
E_0^2/(\omega^2c)$ are a common feature of asymptotic momentum
distributions for ionization in the relativistic regime
\cite{Becker:etal:2002:ATI}, Fig.~\ref{fig:ted1}~(a) reveals also a
shift of the global maximum of the asymptotic momentum distribution from
$(p_{f,x},p_{f,z})=(0,0)$ in the nonrelativistic regime to
$(0,I_p/(3c))$ in the relativistic regime \cite{*[{This is in contrast
    to situations with parameters close to over-the-barrier ionization
    resulting in a different momentum shift, see }] [{}]
  Smeenk:etal:2011:Partitioning_Photon_Momentum,*Titi:Drake:2012:momentum_transfer}.
The shift of the global maximum arises during the under-the-barrier
motion. This may be realized by calculating the distribution of the
kinetic momentum at the exit nonrelativistically and comparing it with
the fully relativistic result; see Fig.~\ref{fig:ted1}~(b).  To identify
the relevant relativistic effects, we expand the Dirac equation
(\ref{eq:DE}) retaining the magnetic dipole correction and the leading
relativistic correction to the electron's kinetic energy (termed
mass-shift correction)
\begin{multline}
  \label{eq:SE}
  \I\partial_t|\Psi\rangle = \\
  \left\{
    \frac{\hat{p}_x^2}{2} + 
    \frac{\hat{p}_y^2}{2} + 
    \frac{\left[\hat{p}_z-{\vec{r}\cdot\vec{E}(t)}/{c}\right]^2}{2} -
    \frac{\hat{\vec p}^4}{8c^2} + 
    V(\vec{r}) + \vec{r}\cdot\vec{E}(t)
  \right\}|\Psi\rangle\,.
\end{multline}
This expansion is justified because relativistic effects for the
under-the-barrier motion are determined by the parameter $I_p/c^2$ and
even for the extreme parameter set used the tunneling process is only
weakly relativistic.  This can be seen from Fig.~\ref{fig:ted1}~(b),
where the calculation based on Eq.~(\ref{eq:SE}) recovers the exact
result from Eq.~(\ref{eq:DE}). Furthermore, the figure shows that the
magnetic dipole term (Lorentz-force) is responsible for the momentum
shift and a decrease of the tunneling probability whereas the mass-shift
correction counteracts the latter effect.

\paragraph{A quasiclassical picture for relativistic tunneling}

For a qualitative interpretation of the tunneling procedure
($\gamma\ll1$) and the momentum shift, we omit the mass-shift correction
as justified above and employ the quasistatic approximation in the
quasiclassical picture. Then the Hamilton function corresponding to
Eq.~(\ref{eq:SE}) yields at maximal laser intensity
\begin{equation}
  \label{eq:Hamiltonian}
  H = \frac{p_x^2}{2} + \frac{p_y^2}{2} + \frac{(p_z+xE_0/c)^2}{2} + 
  V(r)-xE_0 \,,
\end{equation}
with the canonical momenta $p_x$, $p_y$, and $p_z$, $r=\sqrt{x^2 + y^2 +
  z^2}$.  In (\ref{eq:Hamiltonian}) the electric field gauge
\cite{Reiss:1979:gauge,Grynberg:2010:Quantum_Optics} ensures that the
Hamilton function equals the total energy $-I_p$.  Similarly as shown in
the nonrelativistic case \cite{Augst:etal:1989:Tunneling_ionization}, we
estimated numerically that the relativistic deviations still result in
tunneling dynamics in close vicinity along the \mbox{$x$-axis} (the
polarization direction). The atomic potential's central force in the
direction of the \mbox{$y$-axis} and the \mbox{$z$-axis} (the laser's
propagation directions) is then also negligible in the tunneling region
and the canonical momenta $p_y$ and $p_z$ become approximately
conserved.  Consequently, the electron dynamics is separable and the
equation $H=-I_p$ can be written as
\begin{figure}
  \centering
  \includegraphics[width=\linewidth]{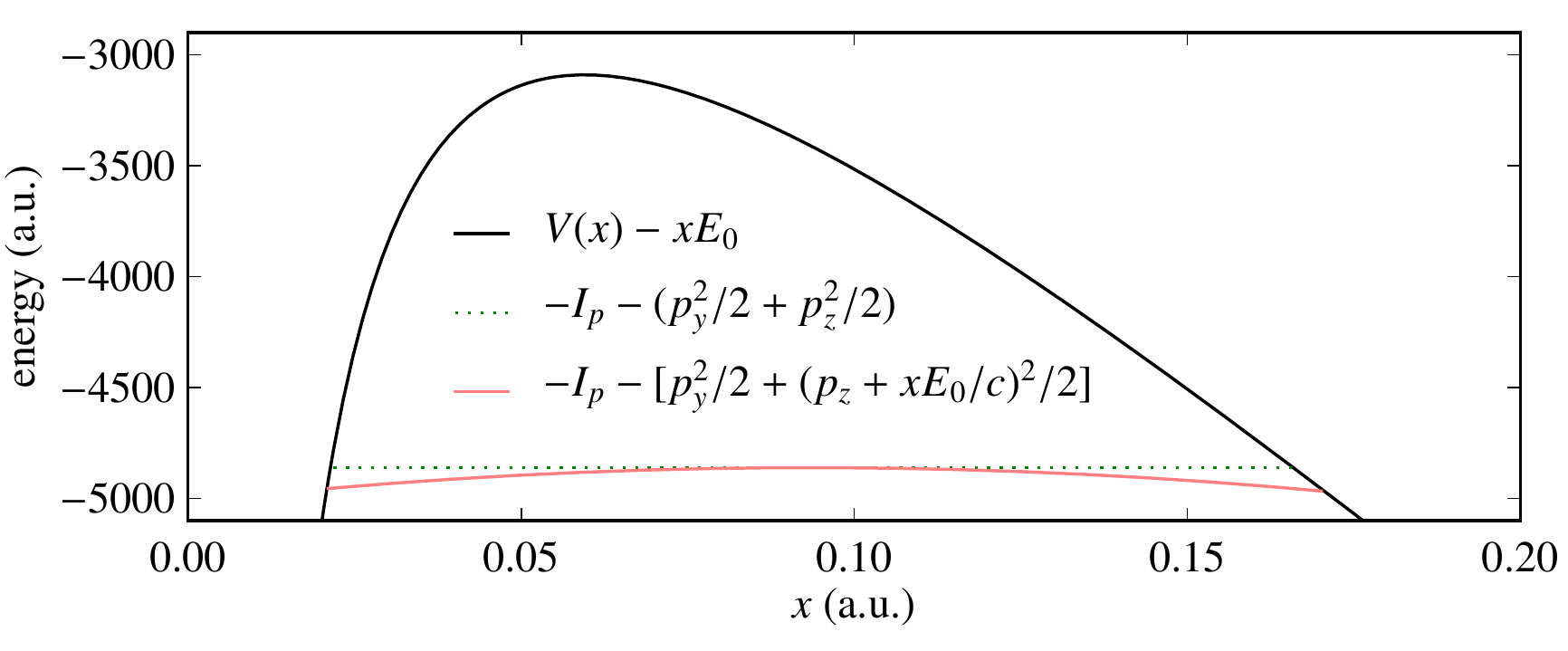}
  \caption{(color online) Scheme of laser induced tunneling in a Coulomb
    potential with $I_p/c^2=0.25$.  The potential barrier $V(x)-xE_0$
    (black line) is displayed with $E_0/E_a=1/30$ and the relativistic
    and nonrelativistic modified energy levels.  The dotted green line
    indicates the description in electric dipole approximation, whereas
    magnetic dipole effects are taken into account for the solid red
    line.  The momenta $p_y$ and $p_z$ are chosen such that the WKB
    tunneling probability is maximal.}
  \label{fig:tpn}
\end{figure}
\begin{equation} 
  \label{eq:1d}
  -I_p-\left[
    \frac{p_y^2}{2} + \frac{(p_z+xE_0/c)^2}{2} 
  \right]= 
    \frac{p_x^2}{2} + \left[V(x)-xE_0\right]\,.
\end{equation}
Equation~(\ref{eq:1d}) is a generalization of an often applied
nonrelativistic one-dimensional model
\cite{Augst:etal:1989:Tunneling_ionization}.  It may be interpreted as
follows.  During the quasi one-dimensional motion along the laser
polarization direction, the electron propagates through the barrier
$V_\text{eff}(x)=V(x)-xE_0$ with a position dependent total energy
$\varepsilon(x)=-I_p-\{p_y^2/2+\mbox{$[p_z+xE_0/c]^2/2$}\}$, which is
the difference between the binding energy and the kinetic energy of the
motion in the propagation direction of the laser.
Equation~(\ref{eq:1d}) yields the momentum in electric field direction
$p_x=\I\sqrt{2[V_\text{eff}(x)-\varepsilon(x)]}$.  Note, that the
kinetic momentum in the propagation direction of the laser
$p_{z,\mathrm{kin}}(x)=p_z+xE_0/c$ in the laser's propagation direction
is shifted due to the Lorentz force during the under-the-barrier motion
causing a modification of $\varepsilon(x)$ and, consequently, a
modification of the momentum $p_x$ as compared to the nonrelativistic
case where the kinetic momentum $p_{z,\mathrm{kin}}(x)=p_z$ is
conserved.  The WKB tunneling probability is proportional to
$\exp(-\Gamma)$, where $\Gamma$ is given by the integral
$\Gamma=-2\I\int_{x_i}^{x_e} p_x\,\mathrm{d}x$ over the classical
forbidden region $x_i\le x \le x_e$
\cite{Popov_1997,*Mur_1998,Milosevic_2002a,*Milosevic_2002d}.  For short
range potentials the WKB tunneling probability is maximal when tunneling
starts with a negative kinetic momentum
$p_{z,\mathrm{kin}}(x_i)=-2I_p/(3c)$ reaching the tunnel exit with
positive $p_{z,\mathrm{kin}}(x_e)=I_p/(3c)$.  At this optimal momentum
the energy $\varepsilon(x)$ is strictly below $-I_p$, which is the
corresponding expression in the case of the electric dipole
approximation, see Fig.~\ref{fig:tpn}. This indicates a decrease of the
maximal tunnel probability with respect to the nonrelativistic case,
cf.~with Fig.~\ref{fig:ted1}~(b).

In the WKB approximation the electron wave function evolves along
quasiclassical trajectories
\cite{Popov_1997,*Mur_1998,Milosevic_2002a,*Milosevic_2002d}. Under the
barrier, they are described by classical Hamiltonian mechanics
developing in imaginary time.  The imaginary time determines the
exponential damping of the ionized electron wave function during
tunneling.  In particular, a time interval $-\I\tau_{K}$ is required to
pass the barrier; thus, the wave function is reduced by a factor
$\e^{-I_p\tau_K}$ while tunneling. Under the barrier the Lorentz force
of order $\upsilon_aB_0/c$ acting for the Keldysh time $\tau_{K}$ causes
a total momentum shift of $\Delta
p_{z,\mathrm{kin}}=p_{z,\mathrm{kin}}(x_e)-p_{z,\mathrm{kin}}(x_i)\sim\tau_K\upsilon_aB_0/c\sim
\mbox{$(x_e-x_i)E_0/c$} \sim I_p/c$.  In order to compensate for the
acceleration in positive \mbox{$z$-direction} by the Lorentz force, the
$z$-component of the initial kinetic momentum at entering the barrier
has to be negative similar to
\cite{Milosevic:etal:2000:hhg,*Walser_2000}.  The kinetic momentum in
the propagation direction of the laser is positive at the barrier exit,
see the dashed blue lines in Fig.~\ref{fig:ted1}~(b), in contrast to the
nonrelativistic case where it vanishes.  Considering the mass-shift
correction in addition to the magnetic dipole terms increases the
tunneling probability; see Fig.~\ref{fig:ted1}~(b).  The general
tunneling picture, however, is not affected.

\paragraph{Tunneling time}

We point out that although the Lorentz force induces a momentum shift
along the laser propagation direction during the under-the-barrier
motion, nevertheless, in the quasistatic WKB approximation the tunneling
process happens instantaneously and thus there is no real coordinate
drift in laser propagation direction as a result of the
under-the-barrier motion.  However, in reality the under-the-barrier
motion requires a small nonzero time span which can be derived by going
beyond the quasistatic WKB theory.

For this purpose we use the exact wave function of the electron in the
quasistatic potential formed by the laser field 
and a zero-range atomic potential. The kinetic energy term 
of the Schr{\"o}dinger equation with the magnetic dipole corrections is
expanded in the coordinate $x$ around the tunnel exit coordinate
$x_e$. Then, the electron wave function in the continuum is derived as
\begin{equation}
  \psi_+(x)=
  T\Airy\left\{
    \mathrm{e}^{-{\I\pi}/{3}}\left[
       2^{1/3}\left(xE_0-I_p\right)/E_0^{2/3}+\phi(x)
    \right]
  \right\}\,,
  \label{eq:psi+}
\end{equation} 
where $\phi(x)=(p_zc+I_p)(3p_zc+2xE_0+I_p)/{3c^2(2E_0)^{2/3}}$, and
$\Airy$ denotes the Airy function. The real transmission coefficient $T$
is obtained by matching $\psi_+(x)$ with the bound state wave function
\cite{Milosevic_2002a,*Milosevic_2002d}
$\psi_0(x)=\mbox{$\sqrt{2\pi\upsilon_a}/[\upsilon_a+(p^2_y+p^2_z)/(2\upsilon_a)]$}
\exp{\{-(\upsilon_a +(p^2_{y}+p_z^2)/(2\upsilon_a)]x\}}$.

\begin{figure}
  \centering
  \hfill
  \includegraphics[scale=0.5]{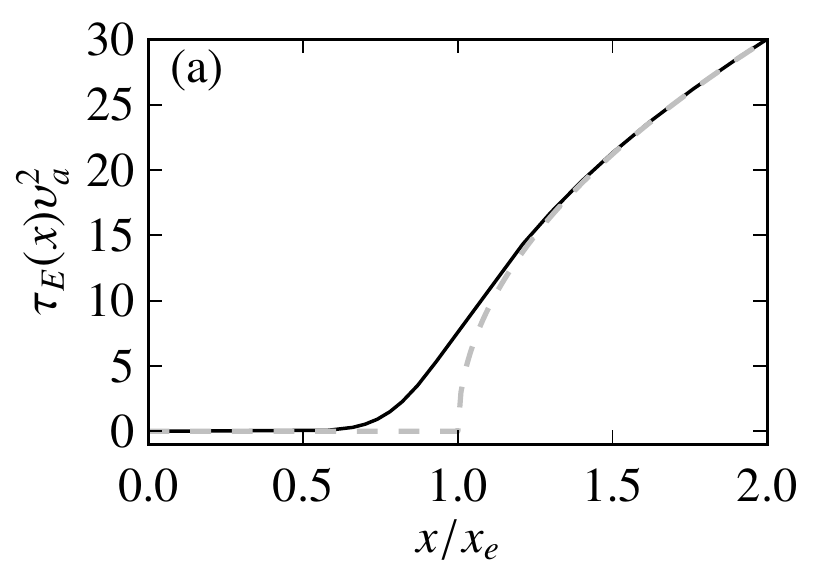}%
  \hfill\hfill
  \includegraphics[scale=0.5]{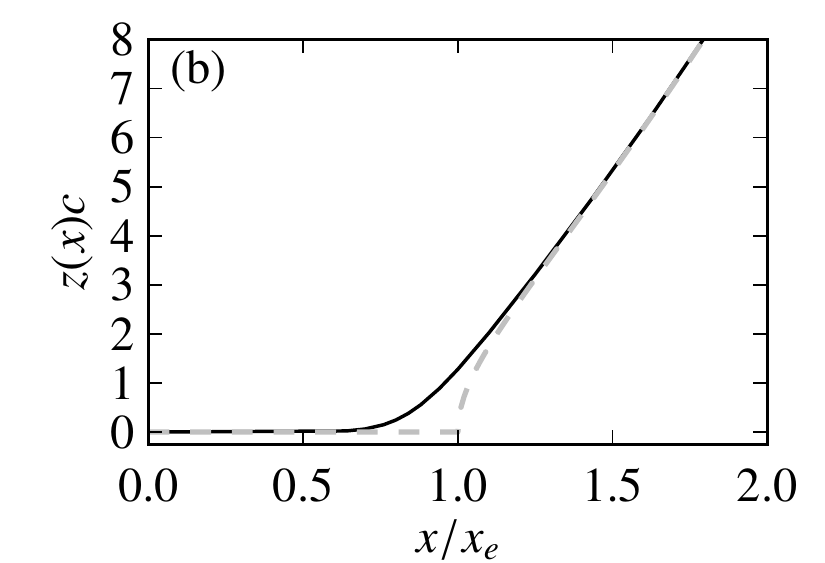}%
  \hfill\mbox{}
  \caption{(a) The scaled time delay vs. the coordinate along the laser
    polarization direction, via a quantum mechanical (black, solid line)
    and a quasiclassical description (gray, dashed line). (b) The
    electron wave packet trajectory in the $x$-$z$-plane, via a quantum
    mechanical (black, solid line) and a quasiclassical description
    (gray, dashed line). In both subfigures $x/x_e<1$ corresponds to
    under-the-barrier motion with $x_e=I_p/E_0$ denoting the tunnel exit
    coordinate. The applied parameters are $I_p/c^2=0.25$,
    $E_0/E_a=1/30$, and the most probable initial kinetic momenta
    $p_{z,\mathrm{kin}}(x_i)=-2I_p/(3c)$ and $p_y=0$.}
  \label{fig:phase}
\end{figure}
The time it takes for the maximum of the wave packet to reach position
$x$ [Eisenbud-Wigner-Smith time delay $\tau_E(x)$] and the drift of the
electron wave packet in propagation direction at position $x$ during the
under-the-barrier motion are determined by the phase of the wave
function $\arg[\psi_+(x)]$ \cite{deCarvalho_2002}, viz.\ %
$\tau_E(x)=-\partial\arg[\psi_+(x)]/\partial I_p
$ and %
$z(x)=-\partial\arg[\psi_+(x)]/\partial p_z
$. Figure~\ref{fig:phase}~(a) shows that the electron spends some time
under the barrier when it is close to the exit.  The tunneling time,
i.\,e., the time delay at the tunneling exit is given by
$\tau_E(x_e)\approx 3^{5/6}\sqrt{\pi}(E_a/E_0)^{2/3}/[2\Gamma(1/6)I_p]$
and the drift distance at the tunneling exit is $z(x_e)
\approx\sqrt{\pi}I_p/[3^{1/6}\Gamma(1/6)E_0^{2/3}c]$.  The tunneling
time of order $1/I_p$ may also be estimated via Heisenberg's uncertainty
relation when assuming an energy uncertainty of order of 
$I_p$.  The drift size along the laser propagation direction at the
tunneling exit is proportional to the tunneling time with a
proportionality factor that is equal to the kinetic momentum in the
laser's propagation direction at the tunnel exit,
$z(x_e)=p_{z,\mathrm{kin}}(x_e)\tau_E(x_e)$. This indicates that the
drift is shaped during the Eisenbud-Wigner-Smith tunneling time. Note
that for the applied parameters the drift size at the tunneling exit
$z(x_e)$ is of order of $1/c$. The drift along the laser propagation
direction during the under-the-barrier motion is due to the Lorentz
force and disappears in the nonrelativistic limit. The time delay
$\tau_E(x)$, however, behaves similarly to that shown in
Fig.~\ref{fig:phase}~(a) also in the nonrelativistic regime.

The time delay far from the tunnel exit, that means
$\mbox{$x-x_e$}\gtrsim E_0^{-1/3}$, follows approximately
$\tau_E(x)|_{x\to \infty}\approx \sqrt{2(x-x_e)/E_0}$.  The latter
corresponds to the quasiclassical motion time of the electron which
tunnels instantaneously and continues at $x_e$ with vanishing initial
momentum in the laser's electric field direction $p_x=0$, reaching $x$
at a moment $t$ ($x-x_e=E_0t^2/2$). Therefore, we can conclude that the
signature of the nonzero tunneling time exists only in a small area near
the tunnel exit at distances $\delta x\sim E_0^{-1/3}$ and disappears at
far distances near the detector, see Fig.~\ref{fig:phase}~(b). This
consequence is relevant also in the nonrelativistic case and explains
the observation of the vanishing tunneling time in the experiments
\cite{Eckle_2008a,*Eckle_2008b,*Pfeiffer_2012}.  Mathematically, the
problem of the vanishing Wigner time at large distances arises when the
laser-induced quasistatic barrier is linear in the x coordinate around
the exit of the tunneling electron in the interval $[x_e-\delta
x,x_e+\delta x]$, which is the case when the saddle point of the
effective potential barrier formed by the laser and atomic potentials is
far from the tunneling exit.  Instead, in the regime close to
over-the-barrier ionization at $E_0/E_a>1/25$ the effective potential
has a parabolic character and analog calculations as carried out here
indicate a measurable Eisenbud-Wigner-Smith tunneling time
\cite{Yakaboylu:etal:2013:unpublished}. In interpreting experimental
results on this tunneling time, the distortion of the electron
trajectories by the Coulomb field of the ionic core during the
sub-barrier as well as out-of-the-barrier motion should be properly
accounted for, see
Ref.~\cite{Popruzhenko:2008:Coulomb-corrected_trajectories,*Yan:Bauer:2012:Sub-barrier_Coulomb_effects,*Baggesen:Madsen:2012:Polarization_Effects,*Kheifets:Ivanov:2010:Delay,*Zhang:Thumm:2010:Electron-ion_interaction,*Nagele:etal:2011:Time-resolved_photoemission,*Ivanov:Smirnova:2011:How_Accurate,*Pazourek:etal:2012:Attosecond_Streaking}.

\paragraph{Numerical simulations}

\begin{figure}
  \centering
  \includegraphics[scale=0.5]{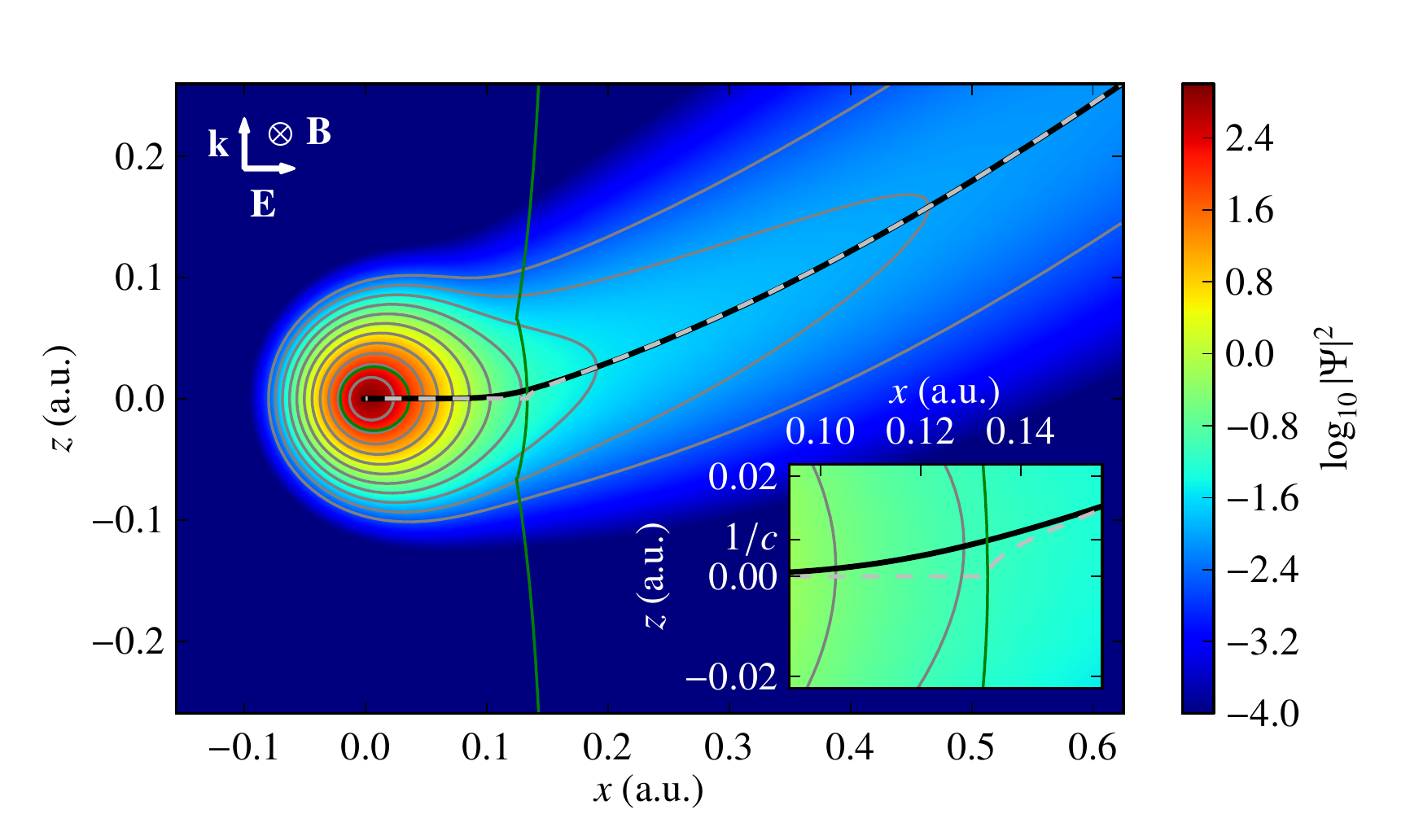}
  \caption{(color online) Numerical simulation of the electron wave
    function in a soft-core potential via the Dirac equation. The
    density plot shows the electron density at the moment when maximal
    laser field strength is attained. The solid black line indicates the
    maximum of the density in the laser propagation direction while the
    dashed line corresponds to the most probable trajectory resulting
    from the quasiclassical description. Solid green lines correspond to
    the border of the classical forbidden region.  White arrows and the
    cross indicate the directions of the laser's electromagnetic fields
    and its propagation direction.  The inset shows a scale-up of the
    region close to the tunnel exit. The applied parameters are
    $I_p/c^2=0.25$, and $E_0/E_a=1/30$, and $\gamma=0.1$.}
  \label{fig:num1}
\end{figure}   

To corroborate our analytical arguments, we have carried out an
\textit{ab initio} simulation of the tunneling process in a highly
charged ion in a laser field of relativistic intensities by solving the
time-dependent Dirac Eq.~(\ref{eq:DE}) numerically
\cite{Bauke:Keitel:2011:Accelerating}.  Figure~\ref{fig:num1} shows the
wave packet of the active electron at the moment when the laser electric
field is at its maximum. Here one part of the wave packet is still bound
or under the barrier while the other part is already in the
continuum. The part under the barrier is not exactly symmetric to the
laser polarization direction. There is a small drift motion into the
laser propagation direction which is of the order $z\sim 1/c$ in
accordance with our estimation above.

\paragraph{Conclusion} 

Laser-driven relativistic tunneling dynamics was shown to be understood
by a simple picture incorporating a scalar tunneling barrier and
position-dependent energy levels. A spatial shift and a momentum shift
along the laser propagation direction at the tunnel exit were identified
as signatures of the relativistic dynamics through the tunneling
barrier. The spatial drift is shown to be proportional to the
Eisenbud-Wigner-Smith time delay, while the momentum shift to the
Keldysh time.  While the spatial shift features intrinsic limitations
for its measurement, the momentum shift is of the order of
10\,a.\,u. for the applied parameters and within the reach of up-to-date
experimental techniques \cite{Eckle_2008a,Eckle_2008b,Pfeiffer_2012}.


We are grateful to C.~M{\"u}ller for valuable discussions at the onset of
the project. 

\bibliography{relti}

\end{document}